\documentclass[showpacs,nofootinbib,preprintnumbers,prd,superscriptaddress,twocolumn]{revtex4}

\usepackage{amsmath}
\usepackage{amsfonts}
\usepackage{amssymb}
\usepackage{graphicx}
\usepackage[titletoc]{appendix}
\usepackage{color}
\usepackage{hyperref}
\usepackage{cleveref}
\usepackage[rightcaption]{sidecap}
\usepackage{subfigure}
\usepackage{comment}

\usepackage{mathtools}

\usepackage{dcolumn}

\usepackage{array}
\usepackage{ctable}
\usepackage{multirow}
\usepackage{siunitx}
\usepackage{longtable}
\usepackage{tabularx}
\usepackage{booktabs}

\graphicspath{{Graphics/}}

\def\be{\begin{equation}}
\def\ee{\end{equation}}
\def\bea{\begin{eqnarray}}
\def\eea{\end{eqnarray}}

\definecolor{vividviolet}{rgb}{0.62, 0.0, 1.0}
\definecolor{amaranth}{rgb}{0.9, 0.17, 0.31}
\definecolor{palatinateblue}{rgb}{0.15, 0.23, 0.89}
\definecolor{brightpink}{rgb}{1.0, 0.0, 0.5}
\definecolor{cornflowerblue}{rgb}{0.39, 0.58, 0.93}
\definecolor{deepcarminepink}{rgb}{0.94, 0.19, 0.22}
\definecolor{radicalred}{rgb}{1.0, 0.21, 0.37}

\hypersetup{ linktoc=all,
    colorlinks, linkcolor={palatinateblue},
    citecolor={brightpink}, urlcolor={amaranth}
}

\begin{document}

\title{Phase-space analysis in non-minimal symmetric-teleparallel dark energy}

\author{Youri Carloni}
\email{youri.carloni@unicam.it}
\affiliation{Universit\`a di Camerino, Via Madonna delle Carceri, Camerino, 62032, Italy.}
\affiliation{INAF - Osservatorio Astronomico di Brera, Milano, Italy.}

\author{Orlando Luongo}
\email{orlando.luongo@unicam.it}
\affiliation{Universit\`a di Camerino, Via Madonna delle Carceri, Camerino, 62032, Italy.}
\affiliation{INAF - Osservatorio Astronomico di Brera, Milano, Italy.}
\affiliation{SUNY Polytechnic Institute, 13502 Utica, New York, USA.}
\affiliation{Istituto Nazionale di Fisica Nucleare (INFN), Sezione di Perugia, Perugia, 06123, Italy.}
\affiliation{Al-Farabi Kazakh National University, Al-Farabi av. 71, 050040 Almaty, Kazakhstan.}

\begin{abstract}
We modify the symmetric-teleparallel dark energy through the addition of a further Yukawa-like term, in which the non-metricity scalar, $Q$, is non-minimally coupled to a scalar field Lagrangian where the phion acts as quintessence, describing dark energy. We investigate regions of stability and find late-time attractors. To do so, we conduct a stability analysis for different types of physical potentials describing dark energy,  namely the power-law, inverse power-law, and exponential potentials. Within these choices, we furthermore single out particular limiting cases, such as the constant, linear and inverse potentials. For all the considered scenarios, regions of stability are calculated in terms of the signs of the coupling constant and the exponent, revealing a clear degeneracy among coefficients necessary to ensure stability. We find that a generic power-law potential with $\alpha > 0$ is not suitable as a non-minimal quintessence potential and we put severe limits on the use of inverse potential, as well. In addition, the equations of state of each potential have been also computed. We find the constant potential seems to be favored than other treatments, since the critical point appears independent of the non-minimal coupling.
\end{abstract}


\maketitle

\section{Introduction}

Exploring the nature of dark energy and dark matter represents a challenge for modern cosmology \cite{Sahni:2004ai,Oks:2021hef,Arbey:2021gdg}. Accordingly, going  beyond general relativity (GR) has recently acquired great importance \cite{Nojiri:2017ncd}. Indeed, although GR successfully passed numerous experimental tests, its status as the ultimate theory of gravitational interaction is often questioned at both infrared and ultraviolet scales. Particularly, the cosmological concordance model, known as the $\Lambda$CDM paradigm, is still plagued by conceptual issues related to the physical interpretation of the cosmological constant, $\Lambda$ \cite{Perivolaropoulos:2021jda} and by recent cosmological tensions \cite{DiValentino:2021izs}.
While various approaches have been developed toward the nature of dark energy and dark matter, including dynamical scalar fields \cite{Peebles:2002gy,Frusciante:2019xia,vandeBruck:2022xbk,Kase:2019veo,Xu:2021xbt}, unified dark energy-dark matter models \cite{Dunsby:2016lkw,Boshkayev:2019qcx,Muccino:2020gqt}, phenomenological scenarios \cite{Copeland:2006wr,Capozziello:2022jbw,Luongo:2015zgq}, etc., a definitive answer to the dynamical problem of the universe is still missing.

On the one hand, Riemannian geometry, i.e., the geometrical structure underlying GR, is interestingly a special case of the more general metric-affine geometry and, specifically, there is no \emph{a priori} need to favor it over other metric-affine descriptions of gravity.

Indeed, the well-known three different, albeit physically equivalent, descriptions of gravity, employing either curvature, torsion or non-metricity, constitute the well-established\footnote{Even though appealing and formally equivalent, the torsion introduces the existence of a further spin, while the non-metricity violates the metric principle. Only GR does not show any further requirement, as it makes use of the curvature.} \emph{geometric trinity of gravity} \cite{BeltranJimenez:2019esp}.

There, the action responsible for describing the universe large-scale structures is constructed through scalar curvature, $R$,  scalar torsion, $T$, or non-metricity scalar, $Q$, respectively. The corresponding theories are known as  GR, teleparallel equivalent of GR and symmetric-teleparallel equivalent of GR.

In this respect, over the recent decades, extended and/or motivated theories of gravity have been investigated to  across various scales \cite{Capozziello:2019cav,Bamba:2012cp}, and in analogy to the gravitational trinity, $f(R)$ \cite{Sotiriou:2008rp,Carloni:2018yoz,Rosa:2019ejh,Goncalves:2023umv}, $f(T)$ \cite{Nojiri:2017ncd,Goncalves:2023umv}, and $f(Q)$ \cite{Heisenberg:2023lru, BeltranJimenez:2017tkd,Koussour:2023rly,Rosa:2023qun,Paliathanasis:2023hqq} scenarios can extend the aforementioned approaches to provide suitable and, mostly alternative, descriptions of dark energy and dark matter, addressing \emph{de facto} limitations of the $\Lambda$CDM model.

Motivated by the class of symmetric-teleparallel theories, we here focus on non-minimal coupled $Q$ gravity. Precisely, in analogy to GR, we consider a generic 0-spin scalar field associated with dark energy, non-minimally coupled to the non-metricity scalar through a Yukawa-like interaction. In this picture, we investigate the corresponding stability properties. To do so, we employ a homogeneous and isotropic universe, formulating our treatment in a spatially-flat Friedmann-Lema$\hat i$tre-Robertson-Walker (FLRW) metric. There, we derive the modified Friedmann equations and explore cosmological dynamics using an autonomous system of first-order differential equations reformulated by virtue of dimensionless variables. Hence, we work out various field potentials, namely singling out the exponential potential, together with power law and inverse power law potentials. We study the critical exponents for each case and we constrain the exponent that permits the critical points to arise. Analogously, we explore the regions of stability and, so, we emphasize the regions in which we expect late-time attractors.

The paper is structured as follows\footnote{Throughout the text, the Lorentzian signature $\left(-+++\right)$ is used, as well as natural units, $8\pi G=c=1$.}. In Sect. \ref{sec:Q gravity with non-minimal coupling}, we introduce our Lagrangian in which we propose a further non-minimal coupling between $Q$ and the dark energy field, under the form of a Yukawa-like potential. The corresponding cosmological features are thus reported. In Sect. \ref{sec:phasespace}, we propose the autonomous system of equations utilized to classify the regions of stability and we classify our solutions focusing on each potential form. Finally stability is developed in Sect. \ref{stability}, while conclusions and perspectives of our work are summarized in Sect. \ref{sec:conclusions}.


\section{Theoretical set up}
\label{sec:Q gravity with non-minimal coupling}

In this section, to investigate non-minimally coupled $Q$ theories, it is convenient to start with $f(Q)$ theories, i.e., introducing analytical functions of $Q$ into the Lagrangian, say
\begin{equation}
    \mathcal{S}=\int d^{4}x\sqrt{-g}\left[-\frac{1}{2}f(Q)+\mathcal{L}_{m}\right].\label{eq:action}
\end{equation}
Here, as stated, $f(Q)$ represents an arbitrary function of $Q$, while $\mathcal{L}_{m}$ denotes the Lagrangian density for matter and, as usual, $g$ stands for the metric determinant \cite{Koussour:2023hgl, Mandal:2020lyq}.

In symmetric-teleparallel theories of gravity, both curvature and torsion vanish, leaving the non-metricity tensor $Q_{\lambda\mu\nu}=\nabla_{\lambda}g_{\mu\nu}$ the only quantity quantifying metric change under teleparellel transport. The traces of $Q_{\lambda\mu\nu}$ are thus defined by
\begin{equation}
Q_{\lambda}=Q_{\lambda\hspace{4mm}\mu}^{\hspace{2mm}\mu},\hspace{6mm}\tilde{Q}_{\lambda}=Q^{\mu}_{\hspace{2mm}\lambda\mu}.
\end{equation}

The non-metricity scalar emerging in Eq.~\eqref{eq:action} is described by  the contraction of the non-metricity tensor with the superpotential tensor $P^{\lambda\mu\nu}$, namely
\begin{equation}
    Q=-Q_{\lambda\mu\nu}P^{\lambda\mu\nu},
\end{equation}
and the components of $P^{\lambda\mu\nu}$ can be expressed explicitly as
\begin{equation}
    P^{\lambda}_{\mu\nu}=-\frac{1}{4}\left(Q^{\lambda}_{\hspace{2mm}\mu\nu}-2Q^{\lambda}_{\left(\mu\hspace{2mm}\nu\right)}+Q^{\lambda}g_{\mu\nu}+\tilde{Q}^{\lambda}g_{\mu\nu}+\delta^{\lambda}_{(\mu} Q_{\nu)}\right).
\end{equation}
In our action, reported in Eq.~\eqref{eq:action}, we also add a further scalar field Lagrangian density, representing a quintessence contribution\footnote{For the sake of clearness, it is theoretically possible to model the interaction of $Q$ with a specific scalar field form that mimes dark matter. Associating our underlying field with dark energy will become evident once the corresponding equation of state is computed.}.

Additionally, we justify the use of the above $f(Q)$ description by by introducing a non-minimal coupling between the non-metricity scalar $Q$ and the scalar field $\phi$. Then, in agreement with GR, our action in Eq.~\eqref{eq:action} can formally be extended by
\begin{equation}
    \mathcal{S}=\int d^{4}x\sqrt{-g}\left[-\frac{1}{2}f(Q)+\mathcal{L}_{m}+\mathcal{L}_{\phi}\right],\label{eq:newaction}
\end{equation}
yielding a $f(Q)$ definition imposed by the ansatz $f(Q)=Q+\xi Q\phi^{2}$, with $\mathcal{L}_{\phi}=-\frac{1}{2}\partial_{\mu}\phi\partial^{\mu}\phi-V\left(\phi\right)$.

In our picture, $\xi$ indicates the coupling constant strength, whereas $V\left(\phi\right)$ represents the the scalar field potential, responsible for the dark energy behavior.

This scenario corresponds to a non-minimal quintessence in $Q$ gravity and it can be denoted as non-minimal symmetric-teleparellel dark energy. This justifies the choice $f(Q)=Q+\xi Q\phi^2$. For the sake of clearness, in fact, the function $f(Q)$ is instead a superfield made up by a double-field approach, say $f(Q,\phi)$. However, by virtue of the additivity of the energy momentum tensor we can naively assume $f(Q)$ as above.

The variation of Eq.~\eqref{eq:newaction} with respect to the metric provides, in fact, the following field equations
\begin{equation}
\begin{split}
    &\frac{2}{\sqrt{-g}}\nabla_{\lambda}\left(\sqrt{-g}f_{Q}P^{\lambda}_{\mu\nu}\right)+\frac{1}{2}g_{\mu\nu}f\\
    &+f_{Q}\left(P_{\mu\lambda\beta}Q_{\nu}^{\lambda\beta}-2Q_{\lambda\beta\mu}P^{\lambda\beta}_{\nu}\right)=T_{\mu\nu}^{(m)}+T_{\mu\nu}^{(\phi)},
\end{split}
\end{equation}
in which the contribution due to quintessence appears in the derivatives of $f(Q)$ and on the right side. In the above relation, we used the standard nomenclature, $f_{Q}=\frac{\partial f(Q)}{\partial Q}$, and the usual definition of the energy-momentum tensor, $T^{(m)}_{\mu\nu}=-\frac{2}{\sqrt{-g}}\frac{\delta\left(\sqrt{-g}\mathcal{L}_{m}\right)}{\delta g_{\mu\nu}}$ and $T^{(\phi)}_{\mu\nu}=-\frac{2}{\sqrt{-g}}\frac{\delta\left(\sqrt{-g}\mathcal{L}_{\phi}\right)}{\delta g_{\mu\nu}}$.

\subsection{Non-minimal $Q$ cosmology}

According to the cosmological principle, we consider the spatially flat FLRW line element, $
    ds^2=-dt^2+a^2(t)\left[dx^2+dy^2+dz^2\right]$, where $a(t)$ is the scale factor.

In FLRW, the non-metricity scalar takes the simple form $Q=6H^2$ and the modified Friedmann equations become \cite {BeltranJimenez:2019tme, Koussour:2023hgl,Khyllep:2022spx}
\begin{subequations}
    \begin{align}
        &3H^{2}=\frac{1}{2f_{Q}}\left(\rho+\frac{f}{2}\right),\\
        &\dot{H}+\left(3H+\frac{\dot{f}_{Q}}{f_{Q}}\right)H=\frac{1}{2f_{Q}}\left(-p+\frac{f}{2}\right),\label{eq:modifiedFried}
    \end{align}
\end{subequations}
where $H\equiv\frac{\dot{a}}{a}$ is the Hubble parameter the dot indicates the derivative with respect to the cosmic time, $t$. The standard Friedmann equations with $f(Q)=Q+\xi Q \phi^2$ read
   \begin{align}
       &3H^{2}=\rho-3H^{2}\xi \phi^{2},\\
        &2\dot{H}+3H^{2}=-p-4\xi \phi \dot{\phi}H-3H^{2}\xi \phi^{2}-2\xi \phi^{2}\dot{H}.
   \label{eq:newmodifiedFried}
\end{align}
Employing the dust case, we have
   \begin{align}
       &H^{2}=\frac{1}{3}\left(\rho^{\phi}_{\rm eff}+\rho_{m}\right),\label{eq:constr}\\
        &2\dot{H}+3H^{2}=-p^{\phi}_{\rm eff}\label{eq:dyn},
   \end{align}
   with the density and pressure acquiring the simple forms,
      \begin{align}
       &\rho^{\phi}_{\rm eff}=\rho_{\phi}-3\xi\phi^{2}H^{2}\label{eq:rho},\\
        &p^{\phi}_{\rm eff}=p_{\phi}+4\xi \phi \dot{\phi}H+3\xi\phi^{2}H^{2}+2\xi\phi^{2}\dot{H}\label{eq:pres}.
        \end{align}
Thus,  combining Eqs.~\eqref{eq:constr} and ~\eqref{eq:dyn}, we obtain the modified Raychaudhuri equation,
\begin{equation}
 \dot{H}=-\frac{1}{2}\left(\rho_{m}+\rho^{\phi}_{\rm eff}+p^{\phi}_{\rm eff}\right),\label{eq:Ray}
\end{equation}
where
       $\rho_{\phi}=\frac{1}{2}\dot{\phi}^{2}+V\left(\phi\right)$ and       $p_{\phi}=\frac{1}{2}\dot{\phi}^{2}-V\left(\phi\right)$,
leading to the following density and pressure for the effective non-minimal case
\begin{subequations}
   \begin{align}
       &\rho^{\phi}_{\rm eff}=\frac{1}{2}\dot{\phi}^{2}+V\left(\phi\right)-3\xi\phi^{2}H^{2},\label{eq:rhoeff}\\
        &p^{\phi}_{\rm eff}=\frac{1}{2}\dot{\phi}^{2}-V\left(\phi\right)+\left(3H^{2}+2\dot{H}\right)\xi \phi^{2} +4\xi \phi \dot{\phi}H.\label{eq:peff}
    \end{align}
\end{subequations}

The dynamical modified Klein-Gordon equation derives from varying the action in Eq.~\eqref{eq:newaction} with respect to the scalar field $\phi$,
\begin{equation}
    \ddot{\phi}+3H\dot{\phi}+V_{,\phi}=-\xi Q\phi,\label{eq:KG}
\end{equation}
in which $\xi Q \phi$ represents a source
term. The usual Klein-Gordon equation is recovered as $\xi \rightarrow 0$.

The results obtained in Eqs.~\eqref{eq:rhoeff},~\eqref{eq:peff} and~\eqref{eq:KG} can be compared to those given in Ref. \cite{Xu:2012jf}.

Selecting a suitable expression for $V\left(\phi\right)$ implies characterizing different stability. Thus, choosing it falls into an autonomous system, allowing for numerical solutions by setting suitable initial values for new proper variables, as we will clarify in the next section.

\section{The phase-space analysis}
\label{sec:phasespace}

Recasting cosmological equations into autonomous systems provides a method to investigate the universe's dynamics \cite{Copeland:2006wr,Capozziello:2022rac,DAgostino:2018ngy,Dimakis:2023cam,Sharma:2021ayk}. We study the cosmological evolution searching for the critical points for a given dark energy potential. Critical points of the autonomous system occur if the derivatives of cosmic variables vanish therein. Specifically, we focus on late times attractors, i.e., those critical points asymptotically behaving as solutions for the autonomous system. To accomplish this, it is convenient to work  new dimensionless variables out, defined as

\begin{subequations}
\begin{align}
&x\equiv \dfrac{\dot{\phi}}{\sqrt{6}H}\,, \hspace{1cm} y\equiv \dfrac{\sqrt{V}}{\sqrt{3}H}\,,\label{eq:var1}\\
& v\equiv \dfrac{\sqrt{\rho_{m}}}{\sqrt{3}H}\,, \hspace{1cm} u\equiv \phi\,.\label{eq:var2}
\end{align}
\end{subequations}

Thus, rewriting the constraint in Eq.~\eqref{eq:constr} and using Eqs.~\eqref{eq:var1}-\eqref{eq:var2}, we obtain
\begin{equation}
x^2+y^2+v^2-\xi u^2=1,\label{eq:constr2}
\end{equation}
whereas the parameter $s=-\frac{\dot{H}}{H^{2}}$ is derived from Eqs.~\eqref{eq:constr}-\eqref{eq:pres} by
\begin{equation}
   s=\frac{3x^{2}+\frac{3}{2}v^{2}+2\sqrt{6}\xi x u}{1+\xi u^{2}}.\label{eq:s}
\end{equation}
At this stage, the dynamical system is given by the conservation equation for $\rho_{\phi}$ in terms of the new variables, say
\begin{equation}
    \begin{cases}
        x'=(s-3)x-\sqrt{6}\xi u-\frac{V_{,\phi}}{\sqrt{6}H^{2}},&\\
        y'=sy+\frac{x}{\sqrt{2}H}\frac{V_{,\phi}}{\sqrt{V}},&\\
        u'=\sqrt{6}x,&
    \end{cases}\label{eq:dynsys}
\end{equation}
where prime indicates  derivative with respect to the number of e-foldings, $N\equiv \ln a$. In addition, the conservation equation for $\rho_{m}$ can be derived by applying the constraint in Eq.~\eqref{eq:constr2}. In this context, we note that the dimensionless energy densities for matter and scalar field can be written as
\begin{equation}
\Omega_{m}=v^{2},\hspace{4mm}\Omega_{\phi}=x^{2}+y^{2}-\xi u^{2},\label{eq:edens}
\end{equation}
respectively. Finally, the dark energy equation of state is denoted as
\begin{equation}
    w_{\phi}=\frac{x^{2}-y^{2}+\xi u^{2}-\frac{2}{3}\xi u^{2} s+4 \sqrt{\frac{2}{3}}\xi x u}{x^{2}+y^{2}-\xi u^{2}}.\label{eq:EoS}
\end{equation}


\subsection{Describing dark energy potentials}

After selecting an appropriate form of $V(\phi)$, Eq.~\eqref{eq:dynsys} transforms into an autonomous system and it can be numerically solved by setting suitable initial values for the set of variables, $\{y,u,v\}$. Appropriate initial conditions can be given as \cite{DAgostino:2018ngy}

\begin{equation}
    y_{i}=10^{-6},\hspace{4mm}
    u_{i}=10^{-6},\hspace{4mm}
    v_{i}^{2}=0.999,
\end{equation}
at $a_{i}=10^{-2}$.

Thus, to find the fixed points and studying the stability around them, we can first impose the condition $x'=y'=u'=0$.

In the following subsection, we explore three distinct forms of scalar field potential for which, imposing the above condition, we are able to argue critical points.


\subsubsection{Power law potential}

The likely most viable form of potential is represented by a power law expression of the form,  $V\left(\phi\right)=V_{0}\phi^{\alpha}$, i.e., a typical potential widely-used for the $\phi$CDM model, as shown in Ref.  \cite{Xu:2021xbt}. In this case, the autonomous system in Eq.~\eqref{eq:dynsys} yields
   \begin{equation}
    \begin{cases}
        x'=(s-3)x-\sqrt{6}\xi u-\alpha\sqrt{\frac{3}{2}}\frac{y^{2}}{u},&\\
        y'=sy+\alpha\sqrt{\frac{3}{2}}\frac{x y}{u},&\\
        u'=\sqrt{6}x.&
    \end{cases}\label{eq:dynsysValpha}
\end{equation}
From the system above, if $\alpha\neq 0$, the critical points are allowed for $\xi<0$ only \footnote{If $\xi>0$, the critical points obtained for $\alpha\neq0$ are not real.}, and those respecting the condition $y\geq 0$ are the following
\begin{align}
    &\left(x,y,u\right)_{I}=\left(0,\sqrt{\frac{2}{2+\alpha}},-\sqrt{\frac{\alpha}{-\xi\left(2+\alpha\right)}}\right),\\
    &\left(x,y,u\right)_{II}=\left(0,\sqrt{\frac{2}{2+\alpha}},\sqrt{\frac{\alpha}{-\xi\left(2+\alpha\right)}}\right).
\end{align}
When substituted into the equation of state, reported in Eq.~\eqref{eq:EoS}, these points yield $w_{\phi}=-1$, \emph{representing a late-time universe with a cosmological constant term as dark energy component}.

Both the critical points exist for $\alpha\geq 0$ and, particularly, we select two particular values, namely  $\alpha=0$ and $\alpha=1$, providing,

\begin{itemize}
    \item[-] a constant potential,
    \item[-] a linear potential,
\end{itemize}
as two main subcases, respectively. In this respect, we single out the two cases above as follows.

\begin{itemize}
    \item[-] {$\alpha=0$ case.} In Ref. \cite{Hrycyna:2015vvs}, exact solutions with $V\left(\phi\right)=V_{0}$ are found for the evolution of a dynamical system within a flat FLRW metric in a universe containing  dust-like matter and a non-minimally coupled scalar field. For $\alpha=0$, the  system in Eq.~\eqref{eq:dynsysValpha}  reduces to
      \begin{equation}
    \begin{cases}
        x'=(s-3)x-\sqrt{6}\xi u,&\\
        y'=sy,&\\
        u'=\sqrt{6}x,&
    \end{cases}\label{eq:dynsysV=const}
\end{equation}
and the critical point reads
    \begin{equation}
        \left(x_{c},y_{c},u_{c}\right)_{I}=\left(0,1,0\right),
    \end{equation}
admitting \emph{a priori} the two cases\footnote{For the sake of completeness, choosing  $\xi$ signs turns out to be of particular interest in field theories applied to early-time cosmology, see e.g. \cite{Belfiglio:2023rxb,Belfiglio:2022yvs,Belfiglio:2022cnd}. }, $\xi>0$ or $\xi<0$.

At this stage, the corresponding normalized energy densities in Eq.~\eqref{eq:edens}  turn into
    \begin{equation}
    \Omega_{m}=0,\hspace{4mm}\Omega_{\phi}=1.
    \end{equation}
    Consequently, for a constant potential, the universe lying on the critical point can be represented by a de Sitter phase, where clearly dark energy behaves as a genuine cosmological constant term\footnote{Here, the issue related to the cosmological constant magnitude at late times is not relevant. For a broad discussion on that, refer to as e.g. Refs. \cite{Luongo:2018lgy,Belfiglio:2023rxb}.}.
\item[-] {$\alpha=1$ case.} This scenario refers to as a symmetric-teleparallel dark energy with a linear potential,  $V\left(\phi\right)=V_{0}\phi$. This potential has been adopted with the aim of alleviating the coincidence problem \cite{Avelino:2004vy}. Hence, through it  Eq.~\eqref{eq:dynsysValpha} becomes
    \begin{equation}
          \begin{cases}
        x'=(s-3)x-\sqrt{6}\xi u-\sqrt{\frac{3}{2}}\frac{y^2}{u},&\\
        y'=sy+\sqrt{\frac{3}{2}}\frac{x y}{u},&\\
        u'=\sqrt{6}x.&
    \end{cases}\label{eq:dynsysV=linear}
    \end{equation}
and, this time, the critical points are given by
    \begin{equation}
        \left(x_{c},y_{c},u_{c}\right)_{I}=\left(0,\sqrt{\frac{2}{3}},-\frac{1}{\sqrt{-3 \xi}}\right),\label{eq:linI}
    \end{equation}
     \begin{equation}
        \left(x_{c},y_{c},u_{c}\right)_{II}=\left(0,\sqrt{\frac{2}{3}},\frac{1}{\sqrt{-3 \xi}}\right).\label{eq:linII}
    \end{equation}
Again, normalized energy densities are determined for both points as
    \begin{equation}
\Omega_{m}=0,\hspace{4mm}\Omega_{\phi}=1,
    \end{equation}
denoting a dark energy dominating the universe at late times.

\end{itemize}
\begin{table*}
\begin{center}
\setlength{\tabcolsep}{0.4em}
\renewcommand{\arraystretch}{2}
\begin{tabular}{c c c c c c c c c}
\hline
\hline
Potential & Critical point & $(x_{c}, \, y_{c}, \, u_{c})$ & $w_{\phi}$  & $\Omega_{m}$ & $\Omega_{\phi}$\\
\hline
$V_{0}$  & I  & $\left(0,\, 1, \, 0\right)$  & $-1$ & $0$ & $1$   \\
\hline
$V_{0}\phi$ & I &  $\left(0,\, \sqrt{\frac{3}{2}}, \,  -\frac{1}{\sqrt{-3 \xi}}\right)$  & $-1$ & $0$ & $1$ \\
& II &  $\left(0,\, \sqrt{\frac{3}{2}}, \,  \frac{1}{\sqrt{-3 \xi}}\right)$  & $-1$ & $0$ & $1$ \\
\hline
$V_{0}\phi^{-1}$ & I &  $\left(0,\, \sqrt{2}, \,  -\frac{1}{\sqrt{\xi}}\right)$  & $-1$ & $0$ & $1$ \\
& II &  $\left(0,\, \sqrt{2}, \,  \frac{1}{\sqrt{\xi}}\right)$  & $-1$ & $0$ & $1$  \\
\hline
$V_{0}e^{-\phi}$ & I  & $\left(0,\, 0, \,  0\right)$ & $\frac{0}{0}$ & 1 & 0 \\
 &  II & $\left(0,\, \sqrt{2\xi-2\sqrt{\xi\left(\xi-1\right)}}, \,  1-\sqrt{\frac{\xi-1}{\xi}}\right)$ & $-1$ & $0$ & $1$\\
 &  III & $\left(0,\, \sqrt{2\xi+2\sqrt{\xi\left(\xi-1\right)}}, \,  1+\sqrt{\frac{\xi-1}{\xi}}\right)$ &  $-1$ & $0$ & $1$\\
\hline
\hline
\end{tabular}
\caption{Results summarizing the different potentials with their critical points. The latter correspond to four distinct models characterized by their corresponding matter and scalar field densities, along with the scalar field equation of state parameter computed \emph{at the critical points}. The choice for the various potentials is motivated throughout the text.}
 \label{tab:critical}
\end{center}
\end{table*}

\subsubsection{Inverse power law potential}

The second framework that we analyze consists in an inverse power law potential, of the form $V\left(\phi\right)=V_{0}\phi^{-\alpha}$ \cite{Xu:2021xbt}. Considering this potential, the dynamical system in Eq.~\eqref{eq:dynsys} assumes the form
   \begin{equation}
    \begin{cases}
        x'=(s-3)x-\sqrt{6}\xi u+\alpha\sqrt{\frac{3}{2}}\frac{y^{2}}{u},&\\
        y'=sy-\alpha\sqrt{\frac{3}{2}}\frac{x y}{u},&\\
        u'=\sqrt{6}x.&
    \end{cases}\label{eq:dynsysV-alpha}
\end{equation}
Even though it is also possible to determine this kind of potential using a power-law as above, with a negative index, we focused on two distinct cases, marking the huge physical differences between the two approaches, see e.g. \cite{Peebles:2002gy}.

The critical points in this system are determined by selecting $\xi>0$ to ensure real solutions and considering $y\geq 0$, we find
\begin{align}
&\left(x_{c},y_{c},u_{c}\right)_{I}=\left(0,\sqrt{\frac{2}{2-\alpha}},-\sqrt{\frac{\alpha}{\xi\left(2-\alpha\right)}}\right),\\
&\left(x_{c},y_{c},u_{c}\right)_{II}=\left(0,\sqrt{\frac{2}{2-\alpha}},\sqrt{\frac{\alpha}{\xi\left(2-\alpha\right)}}\right).
\end{align}

The equation of state in these points is $w_{\phi}=-1$, \emph{giving a cosmological constant at late-times}.

Remarkably, in this case the condition for the existence of critical points limits the exponent to be

\begin{equation}
0\leq \alpha<2,
\end{equation}

where, the case $\alpha=0$ already falls into the previous constant potential. Hereafter, we therefore focus on $\alpha=1$ that naively provides possible viable solutions.

With this value, Eq.~\eqref{eq:dynsysV-alpha} turns into
      \begin{equation}
    \begin{cases}
        x'=(s-3)x-\sqrt{6}\xi u+\sqrt{\frac{3}{2}}\frac{y^{2}}{u},&\\
        y'=sy-\sqrt{\frac{3}{2}}\frac{x y}{u},&\\
        u'=\sqrt{6}x,&
    \end{cases}\label{eq:dynsysV=inverse}
\end{equation}
providing
\begin{align}
&\left(x_{c},y_{c},u_{c}\right)_{I}=\left(0,\sqrt{2},-\frac{1}{\sqrt{\xi}}\right),\\
&\left(x_{c},y_{c},u_{c}\right)_{II}=\left(0,\sqrt{2},\frac{1}{\sqrt{\xi}}\right),
\end{align}
as critical points. Both of them imply
    \begin{equation}
    \Omega_{m}=0,\hspace{4mm}\Omega_{\phi}=1,
    \end{equation}
so we get a universe constituted by dark energy only.

\subsubsection{Exponential potential}

The last potential considered is an exponential potential, i.e., $V\left(\phi\right)=V_{0}e^{-\phi}$. It has been largely investigate in inflation, structure formation and dark energy contexts \cite{Copeland:1997et,Ferreira:1997au}. The choice of this potential give us the following autonomous system
        \begin{equation}
          \begin{cases}
        x'=(s-3)x-\sqrt{6}\xi u+\sqrt{\frac{3}{2}}y^2,&\\
        y'=sy-\sqrt{\frac{3}{2}}x y,&\\
        u'=\sqrt{6}x.&
    \end{cases}\label{eq:dynsysV=exp}
    \end{equation}
By using the exponential potential and the condition $y\geq0$, the critical points obtained are
     \begin{equation}
        \left(x_{c},y_{c},u_{c}\right)_{I}=\left(0,0,0\right),
    \end{equation}
     \begin{equation}
        \left(x_{c},y_{c},u_{c}\right)_{II}=\left(0,\sqrt{2\xi-2\sqrt{\xi\left(\xi-1\right)}}, 1-\sqrt{\frac{\xi-1}{\xi}}\right),
    \end{equation}
      \begin{equation}
        \left(x_{c},y_{c},u_{c}\right)_{III}=\left(0,\sqrt{2\xi+2\sqrt{\xi\left(\xi-1\right)}}, 1+\sqrt{\frac{\xi-1}{\xi}}\right).
    \end{equation}
The existence for the first point is admitted $\forall \xi$, instead for the second point we have to impose $\xi\leq0$ and for the third one $\xi$ need to be $\xi <0$ or $\xi \geq 1$. In the first critical point, the universe is dominated purely by the matter, since
   \begin{equation}
\Omega_{m}=1,\hspace{4mm}\Omega_{\phi}=0,
    \end{equation}
while for the last two points, we obtain  the same de Sitter solution derived with the other potentials, i.e.,
   \begin{equation}
\omega_{\phi}=-1,\hspace{4mm}\Omega_{m}=0,\hspace{4mm}\Omega_{\phi}=1,
    \end{equation}
\emph{indicating again a cosmological constant dominated universe at late-times}.


\section{The stability analysis }
\label{stability}

In this section, we explore the stability of the critical points outlined in \Cref{tab:critical}. The aim is to verify if the derived cosmological solutions can behave as \emph{late time attractors} \cite{Copeland:2006wr,Loo:2023oxk, Ghosh:2023amt, Khyllep:2022spx, Capozziello:2022rac, DAgostino:2018ngy,Dimakis:2023cam,Sharma:2021ayk}. To this end, we evaluate the linear perturbations of the dynamical system and examine the sign of the eigenvalues of the Jacobian matrix associated with each critical point
\begin{eqnarray}
 \label{matJ}
\mathcal{J}=\left( \begin{array}{ccc}
\frac{\partial x'}{\partial x}& \frac{\partial x'}{\partial y} & \frac{\partial x'}{\partial u}\\
\frac{\partial y'}{\partial x}& \frac{\partial y'}{\partial y} & \frac{\partial y'}{\partial u}\\
\frac{\partial u'}{\partial x}& \frac{\partial u'}{\partial y} & \frac{\partial u'}{\partial u}\\
\end{array} \right)_{(x=x_c,y=y_c,u=u_{c})}\,.
\label{eq:J}
\end{eqnarray}
The stability of the solution depends on the eigenvalues, namely as all real parts of the eigenvalues are negative, it establishes a stable point. Conversely, as all are positive, it leads to an unstable point. Interestingly, if the eigenvalue signs are positive and negative, it defines a critical point as a saddle point.

In the first case, the critical point is called attractor. Small perturbation, $\delta x$, $\delta y$ and $\delta u$, around the critical point are obtained as follows
\begin{eqnarray}
\left(
\begin{array}{c}
\delta x' \\
\delta y' \\
\delta u'
\end{array}
\right) = {\mathcal J} \left(
\begin{array}{c}
\delta x \\
\delta y \\
\delta u
\end{array}
\right) \,,
\label{eq:pert}
\end{eqnarray}
where the coefficients of the Jacobian matrix $\mathcal{J}$ depend on the potential under exam.

For each potential, discussed above, we are now in condition to evaluate linear perturbations to search for stability properties.

\subsection{Constant potential}

In the case of constant potential $V\left(\phi\right)=V_{0}$, the matrix $\mathcal{J}$ is
 determined by the following coefficients
    \begin{subequations}
     \begin{align}
&\mathcal{J}_{11}=\dfrac{9x^2+8\sqrt{6}\xi ux-3(1+y^2+\xi u^2)}{2(1+ \xi u^2)},\\
&\mathcal{J}_{12}=-\dfrac{3xy}{1+\xi u^2},\\
&\mathcal{J}_{13}=-\dfrac{\xi\left(3ux^3-3uxy^2+2\sqrt{6}x^2\left(-1+u^2\xi\right)\right)}{(1+\xi u^2)^2}\\
&-\xi\sqrt{6},\\
&\mathcal{J}_{21}=\dfrac{y (3x  + 2 \sqrt{6} u\xi)}{1 + \xi u^2},\\
&\mathcal{J}_{22}=\dfrac{3 (1 + x^2 - 3 y^2) + 3\xi u^2 +  4 \sqrt{6}\xi x u}{2 (1 + \xi u^2)}, \\
&\mathcal{J}_{23}=-\dfrac{y\xi \left(3u(x^2-y^2)+2\sqrt{6}x(-1+\xi u^2)\right)}{(1+\xi u^2)^2}\ , \\
&\mathcal{J}_{31}=\sqrt{6} , \\
&\mathcal{J}_{32}=\mathcal{J}_{33}=0.
    \end{align}
\end{subequations}
Then, the eigenvalue constraint evaluated at critical point $\left(x_{c},y_{c},u_{c}\right)_{I}=\left(0,1,0\right)$ is provided by
\begin{equation}
    \left(3+\mu\right)\left(\mu^2+3\mu+6\xi\right)=0,
\end{equation}
providing the following solutions
\begin{subequations}
    \begin{align}
&\mu_1=-3\ , \\
&\mu_2=\dfrac{1}{2}\left(-3-\sqrt{9-24\xi}\right) , \\
&\mu_3=\dfrac{1}{2}\left(-3+\sqrt{9-24\xi}\right) .
    \end{align}
\end{subequations}

The real parts of eigenvalues are all negative if $\xi >0$. Through this fact, the critical point is stable and indicates a late-time attractor, as shown in \Cref{attractor1}.

\subsection{Linear potential}

The choice of linear potential, $V\left(\phi\right)=V_{0}\phi$, determines the following Jacobian matrix coefficients

\begin{subequations}
    \begin{align}
&\mathcal{J}_{11}=\dfrac{9x^2+8\sqrt{6}\xi ux-3(1+y^2+\xi u^2)}{2(1+\xi u^2)}\ , \\
&\mathcal{J}_{12}=-\dfrac{\sqrt{6}y}{u}-\dfrac{3xy}{1+\xi u^2}\ , \\
&\mathcal{J}_{13}=\dfrac{y^2(\sqrt{6}+2\sqrt{6}\xi u^2+6\xi x u^3+\sqrt{6}\xi^2 u^4)}{2u^2(1+\xi u^2)^2}\, \\
&-\frac{\xi\left(3x^3 u+2\sqrt{6}x^2(-1+\xi u^2)+\sqrt{6}(1+\xi u^2)^2\right)}{(1+\xi u^2)^2}\ , \\
&\mathcal{J}_{21}=\dfrac{y (\sqrt{6} + 6 x u + 5 \sqrt{6}\xi u^2)}{2 u(1 + \xi u^2)}\ , \\
&\mathcal{J}_{22}=\dfrac{\sqrt{6} x + 3 u (1 + x^2 - 3 y^2) + 3\xi u^3 +  5 \sqrt{6}\xi x u^2}{2 u(1 + \xi u^2)}\ , \\
&\mathcal{J}_{23}=\dfrac{y\left(-6 \xi x^2 u^3 + 6\xi y^2 u^3  - \sqrt{6} x (1 - 2\xi u^2+ 5 \xi^2 u^4)\right)}{2u^2(1+\xi u^2)^2}\ , \\
&\mathcal{J}_{31}=\sqrt{6}\ , \\
&\mathcal{J}_{32}=\mathcal{J}_{33}=0\ .
    \end{align}
\end{subequations}

Both the critical points, provided by Eqs.~\eqref{eq:linI} and~\eqref{eq:linII}, yield the same eigenvalues equation,
\begin{equation}
    \left(3+\mu\right)\left(\mu^{2}+3\mu+18\xi\right)=0,
\end{equation}
and, so, once the above equation is solved, the eigenvalues read
\begin{subequations}
    \begin{align}
&\mu_1=-3\ , \\
&\mu_2=-\dfrac{3}{2}\left(1-\sqrt{1-8\xi}\right) , \\
&\mu_3=-\dfrac{3}{2}\left(1+\sqrt{1-8\xi}\right) .
    \end{align}
\end{subequations}
Since for the linear potential the coupling constant is negative, $\xi <0$, then no attractor solutions exist and the critical points are unstable. The same is valid for all the power law potential with $\alpha>0$, indeed for these value not all the eigenvalues are negative\footnote{Remarkably, for a generic exponent, $\alpha>0$, the eigenvalues are easily $\mu_{1}=-3, \mu_{2}=\frac{1}{2}\left(-3-\sqrt{9+24\left(2+\alpha\right)\xi}\right)$ and $\mu_{3}=\frac{1}{2}\left(-3+\sqrt{9+24\left(2+\alpha\right)\xi}\right)$,
and it is evident that not all these values provide negative real parts as $\xi <0$. Thus, the critical points for $\alpha>0$ are not stable and no late-times attractors can be found.}.

\subsection{Inverse potential}

We examine the inverse potential $V\left(\phi\right)=V_{0}\phi^{-1}$ with the following coefficients for matrix $\mathcal{J}$

\begin{subequations}
\begin{align}
   &\mathcal{J}_{11}=\frac{-3 \left(\xi  u^2+y^2+1\right)+8 \sqrt{6} \xi  u x+9 x^2}{2 \left(\xi  u^2+1\right)},\\
   &\mathcal{J}_{12}=\frac{\sqrt{6} y}{u}-\frac{3 x y}{\xi  u^2+1},\\
    &\mathcal{J}_{13}=-\frac{\xi  u^2 \left(\sqrt{6} \left(\xi  u^2+1\right)^2+2 \sqrt{6} x^2 \left(\xi  u^2-1\right)+3 u x^3\right)}{\left(\xi  u^3+u\right)^2}\\
    &+\frac{y^2 \left(\sqrt{6} \xi ^2 u^4-6 \xi  u^3 x+2 \sqrt{6} \xi  u^2+\sqrt{6}\right)}{2 \left(\xi  u^3+u\right)^2},\\
    &\mathcal{J}_{21}=\frac{y \left(2 \sqrt{6} \xi  u+3 x\right)}{\xi  u^2+1}-\frac{\sqrt{\frac{3}{2}} y}{u},\\
    &\mathcal{J}_{22}=\frac{3 \xi  u^3+3 \sqrt{6} \xi  u^2 x+3 u \left(x^2-3 y^2+1\right)+\left(-\sqrt{6}\right) x}{2 \left(\xi  u^3+u\right)},\\
    &\mathcal{J}_{23}=\frac{y \left(-6 \xi  u^3 x^2+6 \xi  u^3 y^2+\sqrt{6} x \left(-3 \xi ^2 u^4+6 \xi  u^2+1\right)\right)}{2 \left(\xi  u^3+u\right)^2}, \\
    &\mathcal{J}_{31}= \sqrt{6}, \\
    &\mathcal{J}_{32}=\mathcal{J}_{33}=0.
\end{align}
\end{subequations}
Again, both the critical points respect the following eigenvalues equation
\begin{equation}
    (\mu+3) \left(\mu^2+3 \mu+6 \xi \right)=0,
\end{equation}
which gives the solutions
\begin{align}
&\mu_1=-3\ , \\
&\mu_2=\dfrac{1}{2}\left(-3-\sqrt{9-24\xi}\right) , \\
&\mu_3=\dfrac{1}{2}\left(-3+\sqrt{9-24\xi}\right).
\end{align}
The values are the same obtained in the constant potential case,  indicating
an attractor, $\forall \xi>0$.

\subsection{Exponential potential}

The last case deals with the analysis of the exponential potential, $V\left(\phi\right)=V_{0}e^{-\phi}$. Here, $\mathcal{J}$ coefficients are
\begin{subequations}
    \begin{align}
&\mathcal{J}_{11}=\dfrac{9 x^2 + 8 \sqrt{6}\xi x u - 3 (1 + y^2 + \xi u^2)}{2(1+\xi u^2)}\ , \\
&\mathcal{J}_{12}=y \left(\sqrt{6} - \dfrac{3x}{1+\xi u^2}\right) , \\
& \mathcal{J}_{13}=-\dfrac{\xi \left(3 x^3 u - 3 x y^2 u + 2 \sqrt{6} x^2 (-1 + \xi u^2 )\right)}{(1 + \xi u^2)^2}\\
&-\xi\sqrt{6}\ , \\
& \mathcal{J}_{21}=\dfrac{y \left(2 \sqrt{6} \xi  u+3 x\right)}{\xi  u^2+1}-\sqrt{\frac{3}{2}} y\ , \\
& \mathcal{J}_{22}= \dfrac{3(1 + x^2 - 3 y^2 + \xi u^2)  - \sqrt{6} x (1 - 4\xi u + \xi u^2)}{2(1+\xi u^2)}\ , \\
& \mathcal{J}_{23}=-\dfrac{\xi y \left(3u(y^2 - x^2) + 2 \sqrt{6} x (-1 + \xi u^2)\right)}{(1 +\xi u^2)^2}\ , \\
&\mathcal{J}_{31}=\sqrt{6}\ , \\
&\mathcal{J}_{32}=\mathcal{J}_{33}=0\ .
    \end{align}
\end{subequations}
In this scenario, three distinct eigenvalues equations occur, one corresponding to each critical point. For the first critical point, we have
\begin{equation}
    \left(3-2\mu\right)\left(2\mu^2+3\mu+12\xi\right)=0,
\end{equation}
whose solutions are
\begin{subequations}
    \begin{align}
&\mu_1^{(I)}=\dfrac{3}{2}\ , \\
&\mu_2^{(I)}= \dfrac{1}{4}\left(-3 - \sqrt{9 - 96 \xi}\right) , \\
&\mu_3^{(I)}=\dfrac{1}{4}\left(-3 + \sqrt{9 - 96 \xi}\right) .
    \end{align}
\end{subequations}

\begin{figure*}
\centering
\subfigure[ Constant potential\label{attractor1}]
{\includegraphics[height=0.46\hsize,clip]{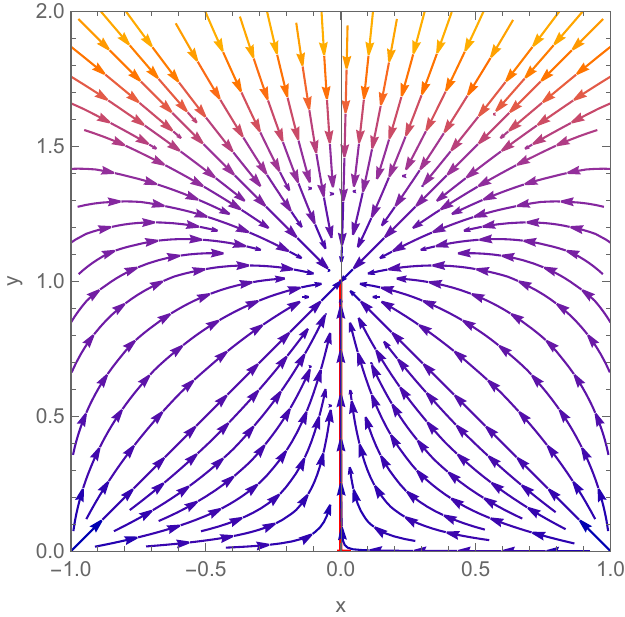}}
\subfigure[ Inverse potential \label{attractor2}]
{\includegraphics[height=0.46\hsize,clip]{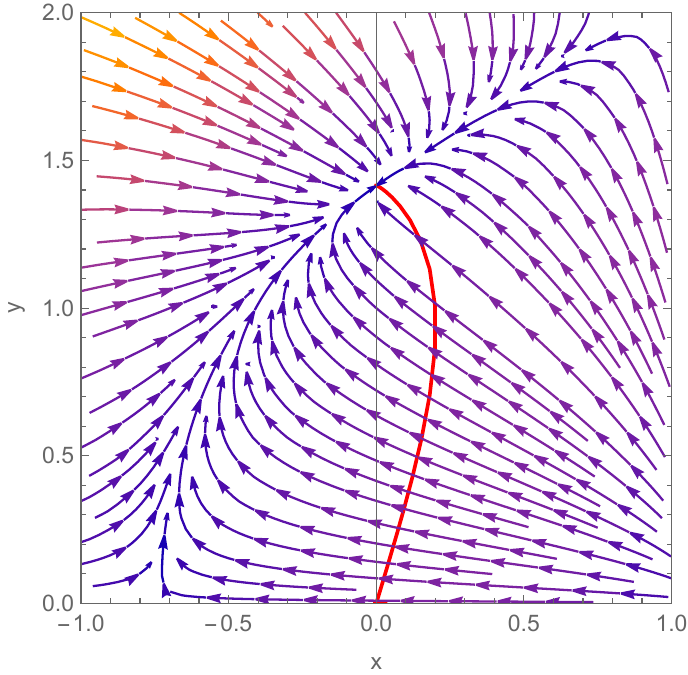}}
\subfigure[ Exponential potential \label{attractor3}]
{\includegraphics[height=0.46\hsize,clip]{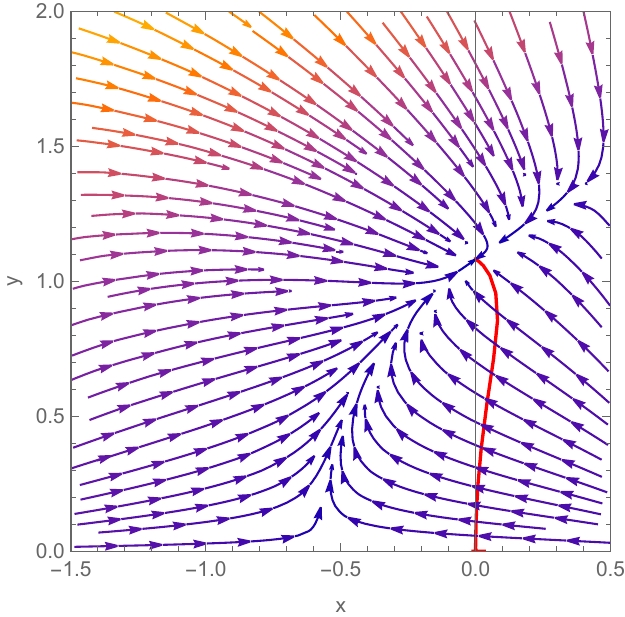}}
\caption{Phase space trajectories on $x-y$ plane for symmetric-teleparallel dark energy. On the top-left, with a constant potential $V(\phi) = V_{0}$ and $\xi = 1$, the critical point $\left(x_{c},y_{c},u_{c}\right)_{I}=\left(0,1,0\right)$ is stable indicating a late-time attractor. On the top-right, considering the inverse potential $V(\phi) = V_{0}\phi^{-1}$ and $\xi = \frac{1}{2}$, the critical points emerging as a late-time attractor are $\left(x_{c}, y_{c}, u_{c}\right)_{I}=\left(0,\sqrt{2},-\sqrt{2}\right)$ and $\left(x_{c}, y_{c}, u_{c}\right)_{II}=\left(0,\sqrt{2},\sqrt{2}\right)$. Finally, on the bottom, the exponential potential case $V(\phi) = V_{0}e^{-\phi}$ is displayed with $\xi = 2$, so the critical point denoting the late-time attractor is $\left(x_{c}, y_{c}, u_{c}\right)_{II}=\left(0,\sqrt{4-2\sqrt{2}},1-\sqrt{\frac{1}{2}}\right)$. The red lines represent the solutions of the dynamical systems.}
\label{PhiandM}
\end{figure*}

These eigenvalues are all positive, so the first critical point is unstable. The second critical point gives us the relation,

\begin{widetext}
\begin{equation}
\dfrac{\xi(3 + \mu)\left[3 \mu + \mu^2 +  6 \left(-2 \xi + \sqrt{\xi(\xi-1)}\right) - 2 \left(-3 \mu - \mu^2 + 6 \xi\right) \left(-\xi + \sqrt{\xi(\xi-1)}\right)\right]}{\left(\xi- \sqrt{\xi(\xi -1)}\right)^2}=0\ ,
\end{equation}

implying as solutions,
\begin{subequations}
    \begin{align}
&\mu_1^{(II)}=-3\ , \\
&\mu_2^{(II)}=-\dfrac{3 - 6 \xi + 6 \sqrt{\xi(\xi-1)} + \sqrt{3} \left[3 + 64 \xi^3 + 4 \sqrt{\xi(\xi-1)} +  8 \xi \left(1 + 5 \sqrt{\xi(\xi-1)}\right) - 8 \xi^2 \left(9 + 8 \sqrt{\xi(\xi-1)}\right)\right]^{1/2}}{2 - 4 \xi + 4 \sqrt{\xi(\xi-1)}}\ , \\
&\mu_3^{(II)}=\dfrac{-3 + 6 \xi - 6 \sqrt{\xi(\xi-1)} + \sqrt{3} \left[3 + 64 \xi^3 + 4 \sqrt{\xi(\xi-1)} +  8 \xi \left(1 + 5 \sqrt{\xi(\xi-1)}\right) - 8 \xi^2 \left(9 + 8 \sqrt{\xi(\xi-1)}\right)\right]^{1/2}}{2 - 4 \xi + 4 \sqrt{\xi(\xi-1)}}\ ,
    \end{align}
\end{subequations}

\end{widetext}

\noindent Hence, for $\xi >1$, all the eigenvalues are negative indicating the stability of the second critical point, acting as an attractor for the universe at late-times, as illustrated in \Cref{attractor2}.

Finally, the third critical point provides the following eigenvalues equation
\begin{widetext}
\begin{equation}
\dfrac{\xi(3 + \mu)\left[-3 \mu - \mu^2 +  6 \left(2 \xi + \sqrt{\xi(\xi-1)}\right) - 2 \left(-3 \mu - \mu^2 + 6 \xi\right) \left(-\xi + \sqrt{\xi(\xi-1)}\right)\right]}{\left(\xi+ \sqrt{\xi(\xi -1)}\right)^2}=0\ ,
\end{equation}
whose solutions are
\begin{subequations}
    \begin{align}
&\mu_1^{(III)}=-3\ , \\
&\mu_2^{(III)}=-\dfrac{-3 + 6 \xi + 6 \sqrt{\xi(\xi-1)} + \sqrt{3} \left[3 + 64 \xi^3 - 4 \sqrt{\xi(\xi-1)} +  8 \xi \left(1 - 5 \sqrt{\xi(\xi-1)}\right) + 8 \xi^2 \left(-9 + 8 \sqrt{\xi(\xi-1)}\right)\right]^{1/2}}{-2 + 4 \xi + 4 \sqrt{\xi(\xi-1)}}\ , \\
&\mu_3^{(III)}=\dfrac{3 - 6 \xi - 6 \sqrt{\xi(\xi-1)} + \sqrt{3} \left[3 + 64 \xi^3 - 4 \sqrt{\xi(\xi-1)} +  8 \xi (1 - 5 \sqrt{\xi(\xi-1)}) + 8 \xi^2 (-9 + 8 \sqrt{\xi(\xi-1)})\right]^{1/2}}{-2 + 4 \xi + 4 \sqrt{\xi(\xi-1)}}\ .
    \end{align}
\end{subequations}

\end{widetext}
It is evident that either for $\xi<0$ or $\xi \geq1$ not all eigenvalues are real and negative, implying the third critical point is unstable.

\section{Outlooks}\label{sec:conclusions}

In this paper, we explored a modified $Q$ theory of gravity that involves a non-minimal coupling between the non-metricity scalar and a dark energy scalar field. We referred to this scenario as non-minimal symmetric-teleparallel dark energy, where the phion  field acts as quintessence non-minimally coupled with gravity, in analogy to the well-established geometric couplings with the Ricci scalar in standard Einstein's gravity.

Within this framework, we searched for regions of stability and, so, to identify late-time attractors, we conducted a stability analysis for different types of potentials. Particular attention has been devoted to the coupling constant sign and strength, showing how it acts to modify the stability of the system, compared with the free Lagrangian, i.e., the one without non-minimal coupling.

Concerning the potentials, among all the possibilities, we specifically considered power-law, inverse power-law, and exponential potentials that appear as viable cases of dark energy potentials.

For power-law potential, we distinguished $\alpha = 0$ and $\alpha = 1$. In the first case, the critical point identifies a late-time attractor since it is real even for $\xi > 0$. However, for $\alpha = 1$, and generally for $\alpha > 0$, $\xi < 0$,  critical points turn out to be unstable. In addition, we conducted a detailed analysis of the linear potential, indicating analogous outcomes.

For inverse power-law potential, we focused on $0 \leq \alpha < 2$ with $\xi > 0$. By choosing $\alpha = 1$ and $\xi = \frac{1}{2}$, both critical points appear stable.

Finally, when analyzing the exponential potential, only $\left(x_{c},y_{c},u_{c}\right)_{II}$ is an attractor solution for the dynamical system, implying $\xi > 1$. Consequently, we ended up that a generic power-law potential with $\alpha > 0$ is not suitable as a non-minimal quintessence potential in the equivalent $Q$ modified gravity scenario.

In addition, the equations of state have been also computed for each case, prompting how they asymptotically tend to a de Sitter case. In this respect, we noticed the evidence for a degeneracy among exponents and coupling constants that emerged as a direct consequence of our analysis, i.e., in order to guarantee stability and physical soundness of our potentials. Specifically, even in our case the constant potential appeared particularly straightforward since the critical point does not depend on the non-minimal coupling constant strength. This output is not surprising, as it corresponds to a case resembling the cosmological constant contribution.

Future works will analyze the use of alternative versions of quintessence, where the scalar field can act differently than stiff matter, as for quintessence. For example, we intend to work fields that can adapt to dark matter instead of dark energy and/or to unified dark energy models. More complicated versions of non-minimal couplings will also be taken into account. Finally, we will investigate more deeply whether in modified theories of gravity, complicated couplings are allowed, possibly providing different information than pure GR.

\begin{acknowledgements}
The authors express their gratitude to Rocco D'Agostino for discussions toward the computational parts of this work. The authors are also grateful to Roberto della Ceca and Luigi Guzzo for their great support during the time spent in INAF-Brera, where this work has been finalized.
\end{acknowledgements}

\bibliographystyle{unsrt}


\end{document}